\documentclass[
journal=jacsat, 
manuscript=article]{achemso}

\usepackage[version=3]{mhchem} 

\author{Alvaro Casas-Bedoya}
\email{casas@physics.usyd.edu.au}
\affiliation{Centre for Ultrahigh bandwidth Devices for Optical Systems (CUDOS), Institute of Photonics and Optical Science (IPOS), School of Physics, University of Sydney, NSW 2006, Australia}
\author{Blair Morrison}
\author{Mattia Pagani}
\author{David Marpaung}
\author{Benjamin J. Eggleton}

\title[\texttt{achemso} demonstration]
{Tunable narrowband microwave photonic filter created by stimulated Brillouin scattering from a Silicon nanowire}

\begin{document}

\begin{abstract}
We demonstrate the first functional signal processing device based on stimulated Brillouin scattering in a silicon nanowire. We use only 1 dB of on-chip SBS gain to create an RF photonic notch filter with 48 dB of suppression, 98 MHz linewidth, and 6 GHz frequency tuning. This device has potential applications in on-chip microwave signal processing and establishes the foundation for the first CMOS-compatible high performance RF photonic filter.
\end{abstract}

\noindent Brillouin scattering is a light-sound interaction process that occurs when photons are scattered from a medium by induced acoustic waves \cite{Boyd1992}. Stimulated Brillouin scattering (SBS) is the strongest nonlinear process and manifest optically in ultra-narrow resonances, which have been harnessed in optical fibers for slow light, sensing, and laser applications \cite{Damzen2003,Lee2008a,Smith1991}. Recently, there has been  strong interest in harnessing SBS in integrated platforms and unlocking functionalities unreachable by other means \cite{Eggleton2013a}. This has been demonstrated in various glasses such as silica or chalcogenides \cite{Pant2011}. Although impressive results were achieved, these devices cannot be monolithically integrated with on-chip electro-optic modulators and photodetectors, which are readily available in silicon, to create a compact tunable filter

Unfortunately, for the CMOS-compatible silicon-on-insulator (SOI) platform, SBS has been elusive. The low elastic mismatch between the silicon core and the silicon dioxide substrate result in weak acoustic confinement, preventing build-up of the SBS process. To remedy this, a recent demonstration employed a hybrid approach to remove the substrate while holding the nanowire with $Si_{3}N_{4}$ \cite{Shin2013}. Although these result harnessed SBS in Si nanowires, they required an extra material and fabrication step, increasing overall complexity.

A recent breakthrough achieved forward propagating SBS (FSBS) in a SOI nanowire \cite{VanLaer2015a} by partially releasing the nanowire from its substrate. Here, they showed that SBS is enhanced at the nanoscale by radiation pressure contributions \cite{VanLaer2015a,Rakich2010,Rakich2012a} and verified that electrostriction (a material property) in combination with radiation pressure (a geometrical property) increases the SBS gain. The amount of SBS gain that was reported, including this geometrical enhancement and novel fabrication methodology, was limited to around 4 dB \cite{VanLaer2015a}, which is hardly usable for conventional signal processing applications.

In this paper we report the first functional device for signal processing based on SBS from a silicon nanowires. We employ a novel cancellation technique \cite{Marpaung2015a,Marpaung2013a,Marpaung2014} to harness this modest SBS gain in silicon, creating a high performance, energy efficient microwave photonic notch filter. We use only 1 dB of on-chip SBS gain to create a cancellation microwave photonic notch filter with 48 dB of suppression, 98 MHz linewidth, and 6 GHz frequency tuning. This demonstration establishes the path towards monolithic integration of high performance SBS microwave photonic filters in a CMOS compatible platform such as SOI.     

\begin{figure}[htbp]
	\centering
	\centerline{\includegraphics[width=10cm]{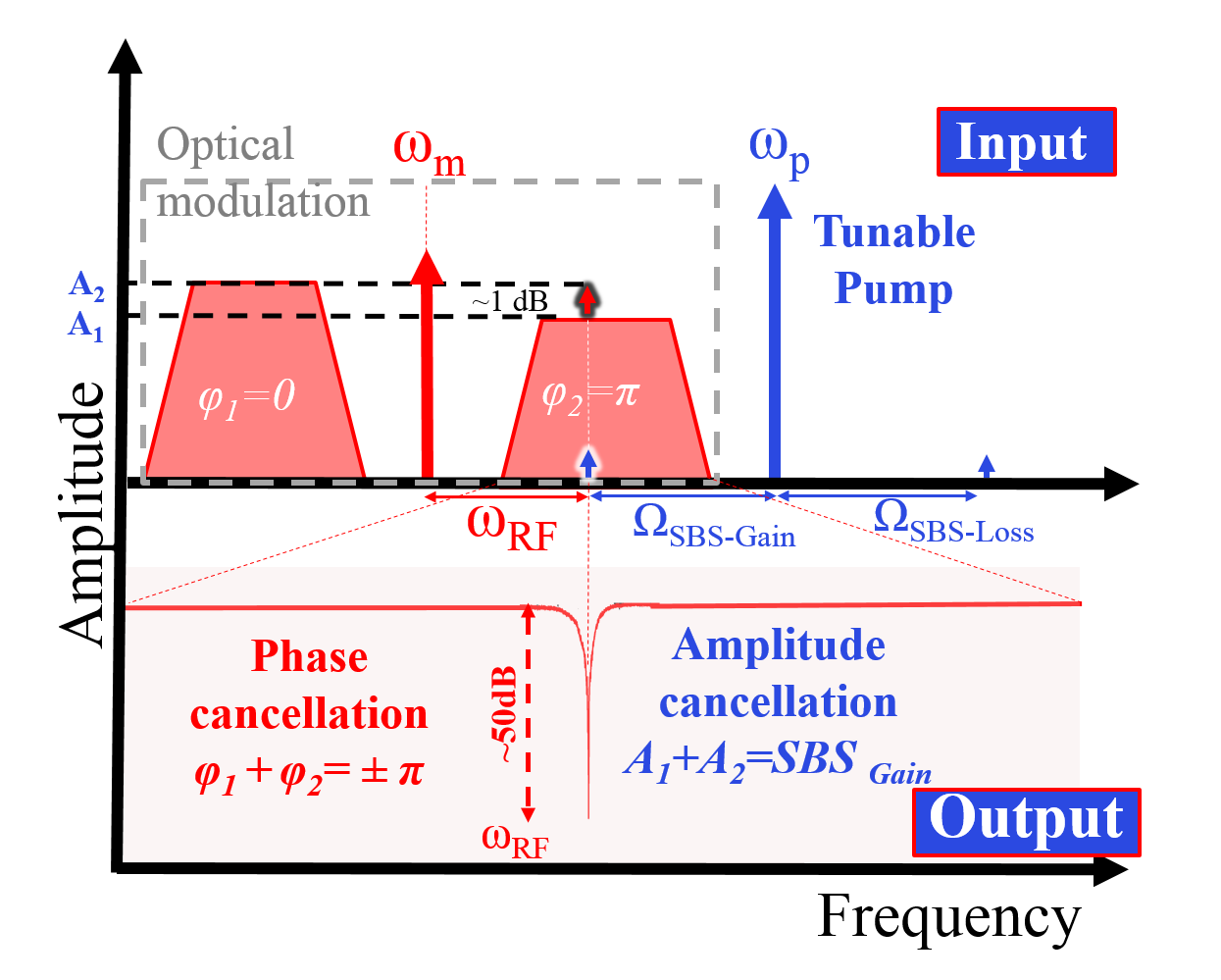}}
	\caption{Radio frequency cancellation technique and working principle of the tunable narrowband microwave photonic filter created by SBS}
	\label{fig1}
\end{figure}

Figure \ref{fig1} summarizes the RF cancellation technique and working principle of the narrowband microwave photonic filter created by SBS. Initially reported in \cite{Marpaung2013a}, the technique uses heterodyne photodetection to generate two RF mixing products between an optical carrier ($\omega_m$) and two modulation sidebands. These two mixing products, having the same frequency, interfere. Destructive interference however occurs only at the frequency ($\omega_{RF}$) where the optical sidebands have the same amplitude (A) and are in antiphase ($\varphi_1 + \varphi_2 =\pm\pi $). This particular amplitude/phase relationship between the sidebands is achieved through SBS ($\varOmega_{SBS Gain}$) on one of the sidebands. Therefore, in a frequency range equal to the SBS linewidth, the RF mixing products interfere destructively, resulting in a very narrowband, high-suppression electrical notch response. The centre frequency of this notch response can be tuned simply by changing position of the SBS resonance on the optical sideband, i.e. tuning the SBS pump wavelength ($\omega_p$).

\begin{figure}[htbp]
	\centering
	\centerline{\includegraphics[width=8  cm]{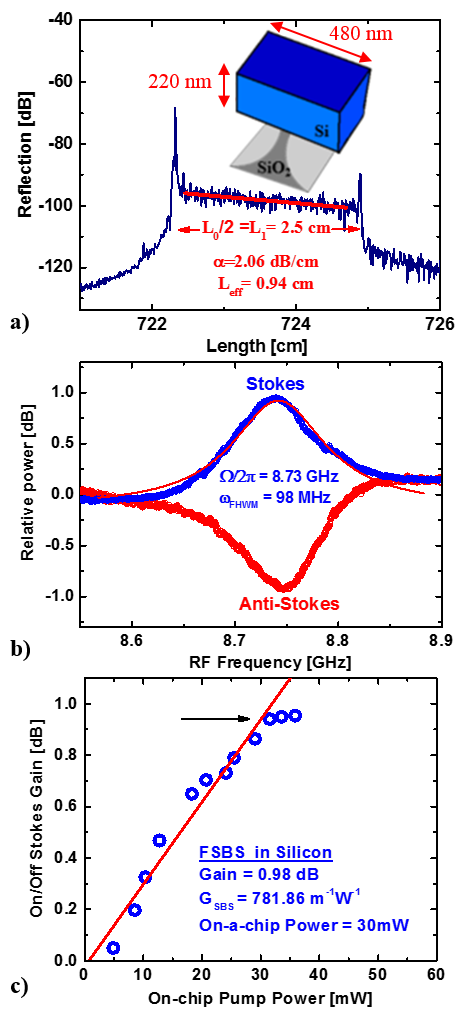}}
	\caption{(a) Optical frequency domain reflectometer measurement of a 1.25 cm SOI nanowire. (Top-figure) Schematic representation of a partially suspended Silicon nanowire (b) Measured Stokes (blue) and anti-Stokes (red) FSBS frequency (c) Maximum Brillouin gain versus input pump power. The arrow indicates the SBS gain saturation at 30 mW due to nonlinear absorption. The red solid line is the fit performed to obtain the Brillouin gain coefficient  ($G_{SBS}$).}
	\label{fig2}
\end{figure}

The SOI nanowire was fabricated at IMEC through ePIXfab. The nanowires were immersed in 10\% diluted hydrofluoric acid with an etching rate of 40 nm min$^{-1}$ for 5.2 minutes to partially release them from the silica substrate. This created a 1.25 cm silicon nanowire with a cross section of 220 by 480 nm which was supported by a silica pillar of 50 nm width [Fig (\ref{fig2}-top)]. As reported in \cite{VanLaer2015a}, such structure restricts the phonon leakage and guarantees high confinement of the optical and acoustic modes allowing an efficient SBS interaction.

We proceeded with the optical characterization of the structure. We coupled 1550 nm  transverse-electric (TE)- polarized light using focus grating couplers \cite{VanLaere2007}  into the chip and measured 5.2 dB coupling efficiency. We used an optical frequency domain reflectometer (OFDR) \cite{LUNA} to study the losses of our nanowire. This methodology infers the time-domain response via Fourier transform from a modulated backscatter signal and consequently allows accurate measurement of the length of the nanowire and propagation losses as shown in Fig \ref{fig2}(a). Here, the first reflection peak is observed 722 cm away from the source. This length corresponds to the total length of our optical fibres just before the chip. The second reflected peak is observed 2.5 cm apart from this first reflection. This value corresponds to twice the length (L) of the nanowire as light is being reflected from the second grating coupler. Propagation losses ($\alpha$) are also obtained with this technique by simply measuring the slope between both grating couplers reflections. This leads to an $\alpha$= 2.06 dB cm$^{-1}$ and an thus an effective length, $L_{eff}$=(1-Exp(-$\alpha$.L))/$\alpha$ = 0.94 cm.

We then measured the SBS gain response of the structure. With 30 mW of coupled pump power, we achieved a 0.98 dB of FSBS gain, which was the highest gain we measured in our structure. The measured Stokes and anti-Stokes have a Lorentzian profile at $\varOmega/{2\pi}$= 8.73 GHz and are combined and plotted in Fig. \ref{fig2}(b). The calculated linewidth was fitted with a Lorentzian curve (red solid line) and calculated to be $\Gamma/{2\pi}$ = 98 MHz. This leads to a quality factor of $Q_{m}$= 89.09 and a phonon life time $\tau$ = 1/$\Gamma$ =1.6 ns.

We measured the maximum stokes gain for different pump powers and plot them in Fig 2 (c).  On resonance, the maximum Brillouin gain as function of Pump power [P$_p$] is:

\begin{equation}
I_s(L) =I_s(0).exp^{G_{SBS}.P_p.L_{eff}},
\label{eq:1}
\end{equation}

where $I_s(L)$ and $I_s(0)$ are the probe intensity respectively at the output and input of the device \cite{Pant2011}. We use equation (1) to infer the SBS gain coefficient ($G_{SBS}$) below nonlinear absorption saturation. Using a linear fitting in Fig. \ref{fig2}(c) we obtain a $G_{SBS}=781.86$ $m^{-1}.W^{-1}$. This relatively low value is due to the phonon leakage trough the post-processed Silica pillar \cite{VanLaer2015a}. However, we emphasise that creating notch filter do not require an ultra-high SBS gain or high pump powers. In fact we show below how using a modest SBS gain and the elegant RF cancellation technique we create an energy- efficiency CMOS-compatible cancellation filter.

\begin{figure}[htbp]
	\centering
	\centerline{\includegraphics[width=10cm]{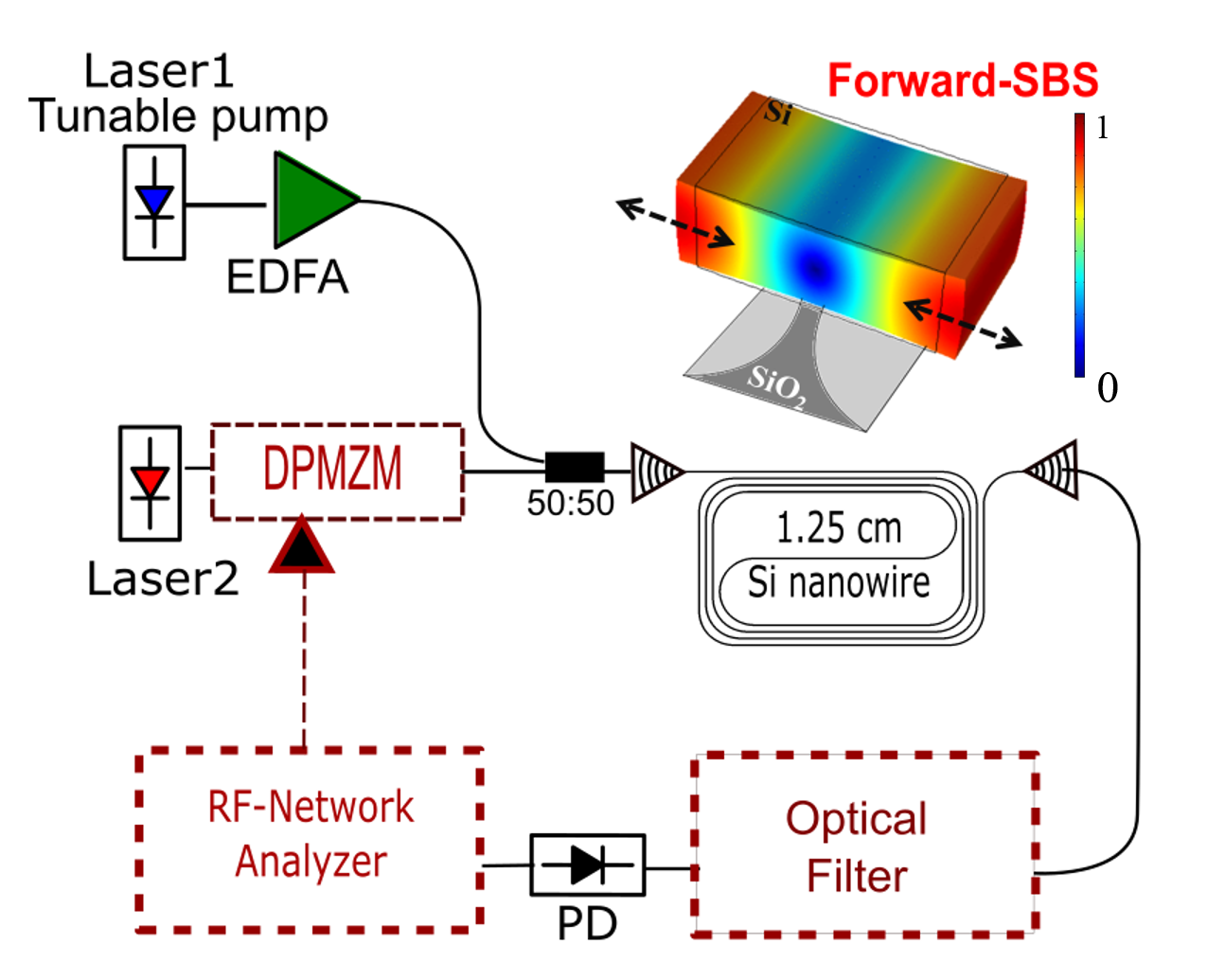}}
	\caption{Set-up of the notch filter experiment.  DPMZM: dual-parallel Mach-Zehnder modulator, PD: photodetector. (Top-right) Simulated transversal acoustic displacement, or forward SBS, from a silicon nanowire}
	\label{fig3}
\end{figure}

\begin{figure}[htbp]
	\centering
	\centerline{\includegraphics[width=10cm]{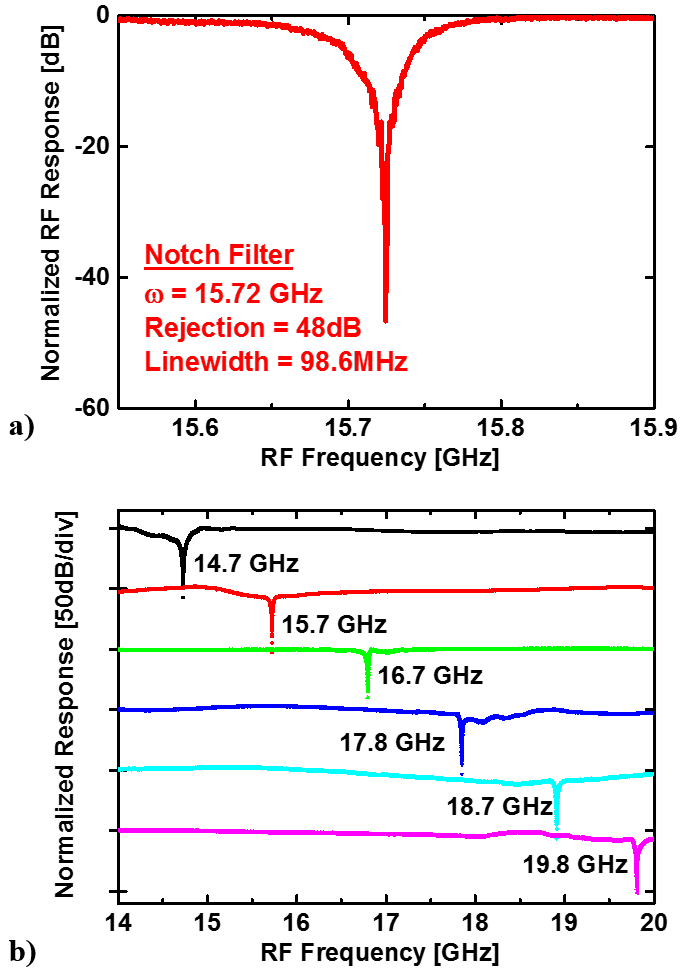}}
	\caption{(a) Measured RF notch filter response at 15.72 GHz (b) Filter frequency tuning where the suppression was kept above 48 dB in all measurements}
	\label{fig4}
\end{figure}

The working principle of the notch filter is described in detail in \cite{Marpaung2015a} and depicted schematically in Fig. \ref{fig3}. Using a dual-parallel Mach-Zehnder modulator (DPMZM), a CW optical carrier is modulated with the input RF signal. The DPMZM bias is set such that the modulation sidebands are $\pi$ out-of-phase, and with amplitude mismatch of 1 dB. The modulated carrier and the SBS pump then co-propagate along a 1.25 cm Si nanowire. The SBS pump frequency and power are set so that a FSBS 0.98 dB gain resonance is induced on the weaker sideband. In this way, only over a 98 MHz frequency range (corresponding to the SBS linewidth), the modulation sidebands have equal amplitude, as well as being $\pi$ out-of-phase. The SBS pump is then removed using a bandpass filter, which selects the modulated carrier and sends it to a photodetector (PD). Upon photodetection, the sidebands and the carrier mix, resulting in the output RF signal. However, in the 98 MHz frequency range where the sideband amplitudes are equal, the mixing products interfere destructively, creating a high-suppression ($>$48 dB) notch response despite the low FSBS gain, as shown in Fig. \ref{fig4}(a).

The centre frequency of the notch in the RF domain can be tuned simply by changing the frequency of the SBS pump. As shown in Fig. \ref{fig4}(b), we were able to continuously tune the notch frequency over a 6 GHz range, while maintaining the notch suppression above 48 dB. To maintain such a high suppression over the entire tuning range is extremely challenging for conventional electronic-based RF notch filters as they exhibit limited resolution (gigahertz instead of megahertz line- widths) and are plagued by trade-offs between key parameters, such as frequency tuning range and suppression. For this reason this demonstration, represents the initial step towards the first tunable integrated megahertz-resolution microwave photonic filter.

We demonstrated the first functional device for signal processing based on SBS in silicon nanowires and establish the crucial first steps towards creation of a high resolution RF signal processor monolithically integrated in a silicon chip, which will be a disruptive technology for next generation radio-frequency systems with wide range of applications.  The filter is continuously tunable in the range of 14 - 20 GHz with a notch suppression of 48 dB and a 3 dB bandwidth of 98 MHz. We believe that combining our results with the already available library of modulators, tunable components, and photodetectors in SOI technologies, our demonstration will establish the foundation towards the first CMOS-compatible high performance RF photonic filter.

This work was funded by the Australian Research Council (ARC) through the Laureate Fellowship (No. FL120100029) Center of Excellence CUDOS (No. CE110001018) and DECRA (DE150101535) programs. We acknowledge support from the U.S. Department of the Air Force through AFOSR/AOARD (grant $\#$ FA2386-14-1-4030). We also acknowledge the fabrication of the SOI nanowires, which were done in the framework of the ePIXnet and the University of New South Wales (UNSW) node of the Australian National Fabrication Facility (ANFF) where the samples were etched

\bibliography{Ref}

\end{document}